\documentclass{article}

\usepackage[nonatbib, preprint]{utils/neurips_2025}

\usepackage[utf8]{inputenc} 
\usepackage[T1]{fontenc}    
\usepackage{hyperref}       
\usepackage{xcolor}         
\definecolor{darkblue}{rgb}{0, 0, 0.85}
\usepackage{soul}

\hypersetup{
    colorlinks = true,
    citecolor = darkblue,
    linkcolor = {black}
}
\usepackage{url}            
\usepackage{booktabs}       
\usepackage{amsfonts}       
\usepackage{nicefrac}       
\usepackage{microtype}      
\usepackage{bm}         
\usepackage{graphicx}		
\usepackage{wrapfig}
\usepackage{amsmath,physics} 
\usepackage{arydshln}
\usepackage{enumitem}
\usepackage{siunitx}
\usepackage{multirow}
\usepackage{hhline}		
\usepackage{xcolor}
\usepackage{mathtools} 
\usepackage{amsthm}			
\usepackage[export]{adjustbox}
\usepackage{algorithm}
\usepackage{algpseudocode}	
\usepackage{xcolor}       
\usepackage{xkcdcolors}
\usepackage{colortbl}
\usepackage{stackengine}
\usepackage{subfigure}
\usepackage[capitalize,noabbrev]{cleveref}
\usepackage[textsize=tiny]{todonotes}
\usepackage{cite} 
\usepackage{pagesel}
\stackMath

\usepackage{wrapfig}
\usepackage{makecell} 
\usepackage{amsmath,amsfonts,bm, xcolor}

\definecolor{lightgreen}{rgb}{.9,1,.9}
\definecolor{trolleygrey}{rgb}{0.5, 0.5, 0.5}
\definecolor{BrickRed}{rgb}{0.6,0,0}
\definecolor{RoyalBlue}{rgb}{0,0,0.8}
\definecolor{Tdgreen}{rgb}{0,0.4,0.7}
\definecolor{pinegreen}{rgb}{0.0, 0.47, 0.44}
\definecolor{cornellred}{rgb}{0.7, 0.11, 0.11}
\definecolor{cadmiumgreen}{rgb}{0.0, 0.42, 0.24}
\definecolor{spirodiscoball}{rgb}{0.06, 0.75, 0.99}
\definecolor{mylightblue}{rgb}{0.85, 0.90, 0.94}
\definecolor{maroon}{cmyk}{0,0.87,0.68,0.32}
\definecolor{mydarkblue}{RGB}{0,0,128}  

\usepackage{soul}
\usepackage{xcolor}

\definecolor{mylightblue}{rgb}{0.85,0.92,1}

\definecolor{lightgreen}{rgb}{.9,1,.9}

\def\eqref#1{equation~\ref{#1}}

\def\1{\bm{1}}

\DeclareMathAlphabet{\mathsfit}{\encodingdefault}{\sfdefault}{m}{sl}
\SetMathAlphabet{\mathsfit}{bold}{\encodingdefault}{\sfdefault}{bx}{n}

\def\etabm{{\bm{\eta}}}

\def\R{\mathbb{R}}
\def\E{\mathbb{E}}

\def\xbm{{\bm{x}}}

\def\zbm{{\bm{z}}}
\def\ybm{{\bm{y}}}

\def\zbm{{\bm{z}}}

\def\mubm{{\bm{\mu}}}

\def\Tsf{{\mathsf{T}}}
\def\Tsf{{\mathsf{T}}}

\def\xbmhat{{\widehat{\bm{x}}}}

\newcommand{\Cov}{\mathrm{Cov}}

\def\defn{\,\coloneqq\,}

\DeclarePairedDelimiterX{\infdivx}[2]{(}{)}{%
  #1\;\delimsize\|\;#2%
}

\usepackage{amsmath,amsfonts,bm}

\theoremstyle{plain}

\newtheorem{theorem}{Theorem}

\usepackage[utf8]{inputenc} 
\usepackage[T1]{fontenc}    
\usepackage{hyperref}       
\usepackage{xcolor}         
\definecolor{darkblue}{rgb}{0, 0, 0.85}
\usepackage{soul}

\hypersetup{
    colorlinks = true,
    citecolor = darkblue,
    linkcolor = {black}
}
\usepackage{url}            
\usepackage{booktabs}       
\usepackage{amsfonts}       
\usepackage{nicefrac}       
\usepackage{microtype}      
\usepackage{bm}         
\usepackage{graphicx}		
\usepackage{wrapfig}
\usepackage{amsmath,physics} 
\usepackage{arydshln}
\usepackage{enumitem}
\usepackage{siunitx}
\usepackage{multirow}
\usepackage{hhline}		
\usepackage{xcolor}
\usepackage{mathtools} 
\usepackage{amsthm}			
\usepackage[export]{adjustbox}
\usepackage{algorithm}
\usepackage{algpseudocode}	
\usepackage{xcolor}       
\usepackage{xkcdcolors}
\usepackage{colortbl}
\usepackage{stackengine}
\usepackage{subfigure}
\usepackage[capitalize,noabbrev]{cleveref}
\usepackage[textsize=tiny]{todonotes}
\usepackage{cite} 
\usepackage{pagesel}
\stackMath

\usepackage{wrapfig}
\usepackage{makecell} 

\title{Moments Matter: Posterior Recovery in Poisson Denoising via Log-Networks}

\author{%
\normalsize Shirin Shoushtari \quad Edward P.~Chandler  \quad Ulugbek S.~Kamilov\\[0.7em]
\small \textnormal{WashU, USA}\\
\footnotesize \texttt{\{s.shirin, e.p.chandler, kamilov\}@wustl.edu}\\[0.5em]
}

\begin{document}

\maketitle

\begin{abstract}
Poisson denoising plays a central role in photon-limited imaging applications such as microscopy, astronomy, and medical imaging. It is common to train deep learning models for denoising using the mean-squared error (MSE) loss, which corresponds to computing the posterior mean $\E[\xbm|\ybm]$. When the noise is Gaussian, the Tweedie’s formula enables approximation of the posterior distribution through its higher-order moments. However, this connection no longer holds for Poisson denoising: while $\E[\xbm|\ybm]$ still minimizes MSE, it fails to capture posterior uncertainty. We propose a new strategy for Poisson denoising based on training a log-network. Instead of predicting the posterior mean $\E[\xbm |\ybm]$, the \textbf{log-network} is trained to learn $\E[\log \xbm |\ybm]$, leveraging the logarithm as a convenient parameterization for the Poisson distribution. We provide a theoretical proof that the proposed log-network enables recovery of higher-order posterior moments and, thus supports posterior approximation.
Experiments on simulated data show that our method matches the denoising performance of standard MMSE models, while providing access to the posterior.
\end{abstract}

\section{Introduction}
\label{sec:intro}
Poisson noise arises naturally in photon-limited imaging applications such as microscopy, astronomy, and medical imaging~\cite{zhang2008wavelets}. Classical approaches to Poisson denoising include variance-stabilizing transforms (VST), which approximate Poisson noise as Gaussian via transformations such as Anscombe or Haar–Fisz, enabling the use of standard Gaussian denoisers~\cite{anscombe1948transformation,zhang2008wavelets,azzari2016variance}. Other direct methods exploit the Poisson likelihood more explicitly, such as total variation regularization~\cite{rudin1992nonlinear} and sparsity-based dictionary learning~\cite{dong2012nonlocal, giryes2014sparsity}, while multi-resolution strategies like PURE-LET further leverage scale-adaptive priors~\cite{luisier2010image, luisier2010fast}. More recently, deep learning methods have been developed for Poisson denoising, ranging from VST-inspired networks~\cite{remez2017deep} to architectures trained directly under Poisson statistics~\cite{lehtinen2018noise2noise,liang2023variational,ta2022poisson2sparse}. These models are typically trained using the mean-squared error (MSE) loss, which yields an approximation of the MMSE estimator.

MMSE denoisers are particularly well understood in the Gaussian setting, where Tweedie’s formula~\cite{Robbins1956Empirical,Miyasawa61, xu2020provable, milanfar2025denoising,kawar2021stochastic} relates the posterior mean $\E[\xbm|\ybm]$ to the score function of the data distribution. Through this connection, the posterior mean implicitly encodes higher-order information that can be recovered via its derivatives~\cite{manorposterior}. In contrast, we show that this elegant property does \emph{not} extend to the Poisson case: although $\E[\xbm|\ybm]$ remains the MMSE estimator, it provides no direct access to the posterior distribution or higher-order moments.

To overcome this limitation, we introduce a new framework for Poisson denoising that trains denoisers directly in the \textit{log-domain}. Specifically, we propose a \emph{log-network} that learns $\E[\log \xbm|\ybm]$, leveraging the fact that the logarithm is the canonical parameterization of the Poisson distribution~\cite{efron2011tweedie,kim2021noise2score}. We show that this design not only preserves denoising accuracy but also opens the door to posterior inference. Our contributions are threefold:
\begin{itemize}
    \item  We provide a theoretical proof that log-domain denoising grants recursive access to higher-order central moments, thereby enabling posterior recovery.
    \item We train a  Poisson denoiser in log-domain and demonstrate that the log-network can estimate higher-order moments and approximate the posterior—capabilities absent in standard MMSE models.
    \item We provide experimental results  demonstrating that the log-network achieves denoising performance on par with standard MMSE models in synthetic Poisson denoising tasks.
\end{itemize}

\begin{figure*}[t]
  \centering
  \includegraphics[width=0.9\textwidth]{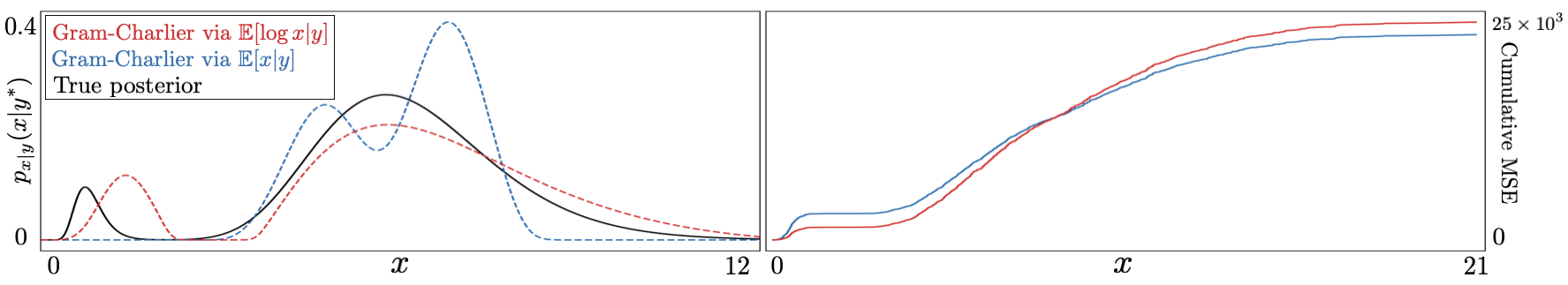}
  \caption{Posterior recovery under a bimodal log-normal prior with observation $y=4$. Left: Posterior recovery comparison. The log-network  better captures the multimodal structure. Right: cumulative MSE across the support, showing lower error for the log-network approach, especially at low signal levels.}
  \label{fig:posterior}

\end{figure*}

\section{Background}
Consider the clean signal $\xbm = [x_1, \cdots, x_n]^\Tsf \in \R^n$. The noisy Poisson observation $\ybm \in \mathbb{N}^n$ is modeled as
\begin{equation} \label{eq:noise_model}
p(\ybm|\xbm) = \prod_{i=1}^n \frac{x_i^{y_i} e^{-x_i}}{y_i!},
\end{equation}
where each $y_i$ is an independent Poisson random variable with mean $x_i >0$. This model naturally arises in photon-limited imaging applications such as microscopy, astronomy, and medical imaging~\cite{zhang2008wavelets}, where photon counts follow Poisson statistics. In practice, acquisition devices are characterized by a gain parameter $\zeta > 0$ that reflects detector sensitivity, leading to the equivalent formulation
\begin{equation}
    \ybm = \zeta \zbm, \qquad z_i \sim \text{Poisson}(x_i / \zeta).
\end{equation}
Smaller values of $\zeta$ correspond to higher noise levels due to reduced photon counts~\cite{le2014unbiased}.

For the exponential family distribution (Gaussian, Poisson, Bernoulli, etc.), there is a general identity relating posterior expectations of the natural (canonical) parameter to derivatives of the log marginal likelihood~\cite{efron2011tweedie}. In the Gaussian case, the natural parameter coincides with the signal itself, and Tweedie’s identity shows that the MMSE denoiser $\E[\xbm| \ybm]$ is directly related to the score function of the marginal distribution. This connection implies that the posterior mean in the Gaussian setting encodes rich structural information, including higher-order moments. In contrast, for the Poisson distribution the natural parameter is $\etabm = \log \xbm$, and Tweedie’s formula applies in this canonical log-domain:
\begin{equation} \label{eq:tweedie}
    \hat{\etabm} = \E[\bm{\eta}|\ybm] = \log{\xbmhat} = \psi (\ybm + 1 ) + \nabla_\ybm \log p_\ybm (\ybm), 
\end{equation}
where $\psi(\cdot)$ is the digamma function and $p(\ybm)$ is the marginal distribution of the observations~\cite{kim2021noise2score}. This formulation highlights that posterior uncertainty in the Poisson setting is naturally structured in log-space, not in the original signal space, motivating the design of denoisers that estimates $\E[\log \xbm|\ybm]$ instead of $\E[\xbm|\ybm]$.

\section{Method}

Our goal is to move beyond standard MMSE denoisers in $x$-space by exploiting the fact that Tweedie’s identity applies in the canonical log-domain in Eq.~(\ref{eq:tweedie}). Using the canonical parameter $\etabm = \log \xbm$, the log-likelihood of the observations is
\begin{equation}
\nonumber \log p(\ybm|\etabm(\xbm)) = \sum_{i=1}^{n} \left( y_i \eta_i - e^{\eta_i} - \log y_i! \right) 
 =  \ybm^\top \etabm - \mathbf{1}^\top \exp(\etabm) - \mathbf{1}^\top \log \ybm!,
\end{equation}
where $!$ denotes the element-wise factorial. The mean estimate of $\etabm$ can be obtained from the mode of $p(\ybm|\etabm(\xbm))$ as:
\begin{equation}
\label{eq:PoissonTweedie_multi}
 \nabla_\ybm \log p(\etabm(\xbm)|\ybm)  =  \nabla_\ybm \log p(\ybm|\etabm(\xbm)) - \nabla_\ybm \log p(\ybm)
 = \etabm(\xbm) - \nabla_\ybm \log \ybm! -  \nabla_y \log p(\ybm) = 0, 
\end{equation}
which implies that 
\begin{equation}\label{eq:TweedieforPoisson_multi}
    \hat{\etabm}(\xbm) = \frac{\nabla \ybm!}{\ybm!} + \frac{\nabla_\ybm p(\ybm)}{p(\ybm)}  = \psi(\ybm+1) + \nabla_\ybm \log p(\ybm),  
\end{equation}
where digamma function $\psi (\cdot)$ is applied element-wise. Thus the MMSE denoiser in the canonical domain is simply the first posterior moment $\mubm_1(\ybm) = \E [\etabm(\xbm)|\ybm]$. The second-order central moment of the posterior is the  covariance matrix $\Cov[\etabm(\xbm)|\ybm]\in \R^{n\times n}$. The $(i_1, i_2)$ entries  of covariance matrix are expressed as
\begin{equation}
\nonumber[\mubm_2(\ybm)]_{i_1,i_2}= \E\left[(\etabm_{i_1}(\xbm)-[\mubm_1(\ybm)]_{i_1})
\big(\etabm_{i_2}(\xbm)-[\mubm_1(\ybm)]_{i_2}\big)
\Big |\ybm\right].
\end{equation}
For any $k \geq 3$, the posterior $k$-th order central moment is a rank-$k$ tensor with entries
\begin{equation*}
[\mubm_k(\ybm)]{i_1,\ldots,i_k}
= \E\left[\prod_{j=1}^k
\big(\etabm_{i_j}(\xbm)-[\mubm_1(\ybm)]_{i_j}\big)
\Big|\ybm\right].
\end{equation*}
The following theorem formalizes how these moments can be recovered directly from the  $\mubm_1(\ybm)$  and its derivatives.

\begin{theorem}\label{thm:thm2}  
In the Poisson model, the posterior central moments of $\etabm = \log \xbm$ satisfy
\begin{align*}
[\mubm_2 (\ybm)]_{i_1, i_2}&=  \frac{\partial [\mubm_1 (\ybm)]_{i_1}}{\partial \ybm_{i_2}}, \\  
[\mubm_3 (\ybm)]_{i_1, i_2, i_3}&=  \frac{\partial [\mubm_2 (\ybm)]_{i_1, i_2}}{\partial \ybm_{i_3}},  \\ 
[\mubm_{k+1}(\ybm)]_{i_1, \cdots,i_k, i_{k+1}} &= \frac{ \partial [\mubm_k(\ybm)]_{i_1, \cdots, i_k}}{\partial \ybm_{i_{k+1}}} + \sum_{j = 1}^{k}   [\mubm_2(\ybm)]_{i_j, i_{k+1}}  [\mubm_{k-1}(\ybm)]_{\ell_j}
\end{align*}
where $ \ell_j \defn \{i_1, \cdots, i_{j-1},  i_{j+1}, \cdots, i_k\}$. This shows that all higher-order moments are uniquely determined by the first posterior moment $\mubm_1(\ybm)$ and its derivatives with respect to $\ybm$.
\end{theorem}

Theorem~\ref{thm:thm2} shows that, unlike in the Gaussian case where higher-order moments can be derived directly from derivatives of $\E[\xbm|\ybm]$, the recursive structure for Poisson noise emerges only in the canonical log-domain. Practically, this means that once a network is trained to approximate $\E[\log \xbm | \ybm]$, its Jacobian with respect to the input yields the posterior covariance, and higher-order derivatives give access to higher-order moments. In this way, the full posterior structure can be systematically recovered. This highlights a key advantage of log-domain denoisers over traditional MMSE denoisers in $x$-space.

\begin{figure*}[t]
  \centering
  \includegraphics[width=\textwidth]{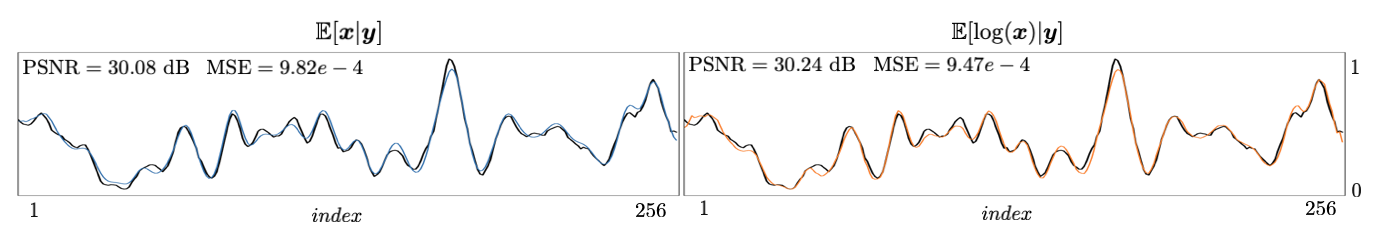}

  \caption{Comparison of denoising performance. Left: standard MMSE network$\E[
\xbm|\ybm]$. Right: proposed log-network $\E[\log \xbm|\ybm]$. Both models achieve similar reconstruction quality (PSNR and MSE), demonstrating that log-domain training maintains competitive denoising accuracy.}
  \label{fig:denoising}
\end{figure*}

\begin{table}[t]
    \centering
    \caption{Quantitative denoising performance. PSNR (dB) and MSE for MMSE denoising $\E[\xbm|\ybm]$ and log-network $\E[\log \xbm|\ybm]$ across different $\zeta$.}
\begin{tabular}{lcccc}
\toprule
& \multicolumn{2}{c}{$\E[\log \xbm | \ybm]$} & \multicolumn{2}{c}{$\E[ \xbm|\ybm]$} \\
\cmidrule(r){2-3} \cmidrule(l){4-5}
$\zeta$ & PSNR & MSE & PSNR & MSE \\
\midrule
16  & 24.12 &0.0038&  24.17&0.0038  \\
32  & 25.92 & 0.0026 & 25.91 & 0.0026 \\
64  & 28.08 & 0.0016 & 28.32 & 0.0015 \\
\bottomrule
\end{tabular}
\label{tab:denoising}

\end{table}


\section{Experiments}
To evaluate posterior recovery, we first construct a toy experiment with a bimodal log-normal prior over  $x \in [0.01,20]$ and a Poisson likelihood. For a fixed observation  $y=4$ we compute the true posterior and its moments, and train two small MLPs for  $\E[x|y]$ and $\E[\log x|y]$. Posterior approximations are then reconstructed via Gram–Charlier expansion~\cite{cramer1999mathematical} from the higher-order moments according to Theorem~\ref{thm:thm2}. As shown in Fig.~\ref{fig:posterior}, the MMSE-based expansion fails to capture the multimodal posterior, while the log-network successfully recovers both modes and achieves lower cumulative MSE, particularly at low intensities where Poisson noise is strongest.

We next compare the two models on a 1D Poisson denoising benchmark. Clean signals $\xbm\in\mathbb{R}^{256}$ are generated from $\mathrm{Gamma} (\alpha=1.5,\beta=2.0)$, smoothed with a Gaussian kernel, normalized to $[0,1]$, and corrupted by Poisson noise at varying gains $\zeta$. Both models use  lightweight 1D architectures with five convolutional layers (kernel size $7$, ReLU/LeakyReLU activations) and a final $1\times 1$ convolution for output prediction. Training is performed with MSE loss on synthetic data generated on-the-fly, with distinct train/validation/test splits and early stopping by validation PSNR. Fig.~\ref{fig:denoising} shows denoising results for $\zeta=64$, where both models achieve nearly identical reconstructions. Table~\ref{tab:denoising} further confirms that PSNR and MSE remain comparable across noise levels. These results highlight that log-domain training preserves denoising accuracy while uniquely enabling posterior recovery.

\section{Conclusion}
We introduced a log-domain training strategy  for Poisson denoising that moves beyond conventional MMSE models. By training denoisers to estimate $\E[\log \xbm |\ybm]$, we showed both theoretically and empirically that the log-network provides recursive access to higher-order posterior moments, enabling posterior recovery. Experiments on synthetic benchmarks confirm that the log-network achieves denoising performance comparable to standard MMSE models while uniquely supporting posterior estimation. This work highlights the value of canonical parameterization in the design of deep denoisers and opens avenues for uncertainty-aware methods in photon-limited imaging.

\bibliographystyle{utils/IEEEbib}
\bibliography{utils/references}
\newpage
\section{Appendix}
\subsection{Proof of Theorem~\ref{thm:thm2}}
\begin{proof}
    $\mubm_1(\ybm)$ can be written as 
\begin{equation}
    \mubm_1(\ybm)= \E [\etabm | \ybm ]  = \frac{\int_{\R^n} \etabm(\xbm) p_{\ybm|\xbm}(\ybm| \xbm) p_\xbm(\xbm) d\xbm }{p_\ybm(\ybm)}. 
\end{equation}
Jacobian of $\mubm_1(\ybm)$ can be written as 
\begin{equation}\label{eq:posteriorder_multi}
 \nabla_\ybm \E [\etabm | \ybm ]  =   \frac{\int_{\R^n} \etabm(\xbm) \left (\nabla_\ybm p_{\ybm|\xbm}(\ybm| \xbm)\right )^\Tsf p_\xbm(\xbm) d\xbm }{p_\ybm(\ybm)} 
   - \frac{\int_{\R^n} \etabm(\xbm) p_{\ybm|\xbm}(\ybm| \xbm) p_\xbm(\xbm) d\xbm }{p_\ybm(\ybm)}.\frac{\left (\nabla p_\ybm(\ybm)\right )^\Tsf}{p_\ybm(\ybm)}. 
\end{equation}
Gradient of $p_{\ybm|\xbm}$ can be calculated using logarithmic derivative as  
\begin{equation}\label{eq:deroflikelihood_multi}
    \nabla_\ybm p_{\ybm|\xbm}(\ybm| \xbm) = p_{\ybm|\xbm}(\ybm| \xbm) \Big ( \log \xbm - \frac{\nabla \ybm!} {\ybm!}\Big ) = p_{\ybm|\xbm}(\ybm|\xbm) \Big ( \etabm(\xbm) - \psi(\ybm+1)\Big ), 
\end{equation}
where we used $ \psi (\ybm+1 ) \defn \nabla \ybm!/\ybm!$ and $\etabm(\xbm)= \log \xbm$. Plugging Eq.~(\ref{eq:deroflikelihood_multi}) into Eq.~(\ref{eq:posteriorder_multi}) and using the Tweedie's formula yields 
\begin{align*}
    \nonumber \nabla_\ybm \E [\etabm | \ybm ] &= 
     \E [\etabm(\xbm)\etabm(\xbm)^\Tsf | \ybm ] - \E [\etabm(\xbm) | \ybm ]  \psi(\ybm+1)^\Tsf  - \E [\etabm(\xbm) | \ybm ]  \left(\E [\etabm(\xbm) | \ybm ]  - \psi(\ybm+1) \right)^\Tsf\\
    \nonumber& = \E [\etabm(\xbm)\etabm(\xbm)^\Tsf | \ybm ]  - \E [\etabm(\xbm) | \ybm ] \E[\etabm(\xbm) | \ybm ] ^\Tsf =\mubm_2(\ybm)
\end{align*}
which completes the proof for $k=1$. For $k\geq 2$, we have 
\begin{align*}
\nonumber[\mubm_k(\ybm)]_{i_1,\ldots,i_k}&= \E\!\left[\prod_{j=1}^k \big(\etabm_{i_j}(\xbm)-[\mubm_1(\ybm)]_{i_j}\big)\;\middle|\;\ybm\right] \\
&= \frac{\int_{\R^n} \prod_{j=1}^k \big(\etabm_{i_j}(\xbm)-[\mubm_1(\ybm)]_{i_j}\big)\,
   p_{\ybm|\xbm}(\ybm|\xbm)\,p_\xbm(\xbm)\,d\xbm}{p_\ybm(\ybm)}. \nonumber
\end{align*}
For any $i_{k+1} \in \{1, \cdots , n\}$, the derivative of $[\mubm_k(\ybm)]_{i_1, \cdots, i_k}$ with respect to $\ybm_{i_{k+1}}$ (denoted as $\nabla_{k+1}$) can be expressed as 
\begin{align}\label{eq:moment_derivative}
\nonumber\frac{\partial [\mubm_k(\ybm)]_{i_1,\ldots,i_k}}{\partial \ybm_{i_{k+1}}}&= \nabla_{k+1}\,
   \E\!\left[\prod_{j=1}^k \big(\etabm_{i_j}(\xbm)-[\mubm_1(\ybm)]_{i_j}\big)\;\middle|\;\ybm\right] \\
&= \frac{\int_{\R^n} \nabla_{k+1}\!\Big(\prod_{j=1}^k (\etabm_{i_j}(\xbm)-[\mubm_1(\ybm)]_{i_j})\Big)\,
   p_{\ybm|\xbm}\,p_\xbm\,d\xbm}{p_\ybm(\ybm)} \nonumber \\
&\quad + \frac{\int_{\R^n} \prod_{j=1}^k (\etabm_{i_j}(\xbm)-[\mubm_1(\ybm)]_{i_j})\,\nabla_{y_{k+1}} p_{\ybm|\xbm}\,p_\xbm\,d\xbm}{p_\ybm(\ybm)} \nonumber \\
&\quad - [\mubm_k(\ybm)]_{i_1,\ldots,i_k}\,\frac{\nabla_{{k+1}}p_\ybm(\ybm)}{p_\ybm(\ybm)}.
\end{align}
We investigate the cases of $k=2$ and $k\geq3$ separately. When $k=2$, the first term reduces to $-\nabla_{\ybm_{i_3}}[\mubm_1(\ybm)]_{i_1}  \E[\etabm_{i_2}(\xbm) -[\mubm_1(\ybm)]_{i_2}|\ybm]-\nabla_{\ybm_{i_3}}[\mubm_2(\ybm)]_{i_2}  \E[\etabm_{i_1}(\xbm) -[\mubm_1(\ybm)]_{i_1}|\ybm] =0$.
In the second term, we use 
$\nabla_{y_{i_3}} p(\ybm|\xbm) = ( \etabm_{i_3}(\xbm) - \psi(\ybm_{i_3}+1) ) p_{\ybm|\xbm}$ from Eq.~(\ref{eq:deroflikelihood_multi}). In the third term, we use Tweedie's formula in Eq.~(\ref{eq:TweedieforPoisson_multi}).
Thus, we have: 
\begin{align}
   \nonumber &\frac{ \partial [\mubm_2(\ybm)]_{i_1, i_2}}{\partial \ybm_{i_3}} 
   = \frac{\int_{\R^n}  \Big(\prod_{j=1}^3 (\etabm_{i_j}(\xbm)-[\mubm_1(\ybm)]_{i_j})\Big)   p_{\ybm|\xbm} p_\xbm d\xbm }{p_\ybm(\ybm)} \\
   \nonumber& + \frac{\int  (\prod_{j=1}^2 (\etabm_{i_j}(\xbm)-[\mubm_1(\ybm)]_{i_j}) )([\mubm_1(\ybm)]_{i_3}-\psi(\ybm_{i_3}+1))  p_{\ybm|\xbm} p_\xbm d\xbm }{p_\ybm(\ybm)}\\
    \nonumber&-[\mubm_2(\ybm)]_{i_1, i_2} \Big ( \etabm_{i_3}(\xbm) - \psi(\ybm_{i_3}+1) \Big )\\ 
    \nonumber & = [\mubm_3(\ybm)]_{i_1, i_2, i_3} + [\mubm_2(\ybm)]_{i_1, i_2} \Big ( \etabm_{i_3}(\xbm) - \psi(\ybm_{i_3}+1) \Big )  \\
    \nonumber&-[\mubm_2(\ybm)]_{i_1, i_2} \Big ( \etabm_{i_3}(\xbm) - \psi(\ybm_{i_3}+1) \Big ) = [\mubm_3(\ybm)]_{i_1, i_2, i_3} ,
\end{align}
where in the last equality, we used the fact that $\etabm_{i_3}(\xbm) = [\mubm_1(\ybm)]_{i_3} $. 
Similarly for $k\geq 3$, the first term can be written as 
\begin{equation}\label{eq:res1_compact}
-\sum_{j=1}^{k}\,
\frac{\partial [\mubm_1(\ybm)]_{i_j}}{\partial \ybm_{i_{k+1}}}\,
\E\!\left[\!\!\,\prod_{\ell\neq j}\big(\etabm_{i_\ell}(\xbm)-[\mubm_1(\ybm)]_{i_\ell}\big)\,\middle|\,\ybm\right] \nonumber = -\sum_{j=1}^{k}
\frac{\partial [\mubm_1(\ybm)]_{i_j}}{\partial \ybm_{i_{k+1}}}\,
[\mubm_{k-1}(\ybm)]_{\ell_j},
\end{equation}
where $ \ell_j \defn \{i_1, \cdots, i_{j-1},  i_{j+1}, \cdots, i_k\}$. For the second term, we obtain the gradient of $p_{\ybm|\xbm}$ using Eq.~(\ref{eq:deroflikelihood_multi}).
Thus, the second term is replaced with: 
\begin{align}\label{eq:res2_compact}
 &\E\!\left[\!\prod_{j=1}^{k+1}(\etabm_{i_j}(\xbm)-[\mubm_1(\ybm)]_{i_j})\,\middle|\,\ybm\right]\nonumber\\
&+ \big([\mubm_1(\ybm)]_{i_{k+1}}-\psi(\ybm_{i_{k+1}}+1)\big)\,
   \E\!\left[\!\prod_{j=1}^{k}(\etabm_{i_j}(\xbm)-[\mubm_1(\ybm)]_{i_j})\right] \nonumber\\
&= [\mubm_{k+1}(\ybm)]_{i_1,\ldots,i_{k+1}} \nonumber\\
 &+ \big([\mubm_1(\ybm)]_{i_{k+1}}-\psi(\ybm_{i_{k+1}}+1)\big)\,
   [\mubm_{k}(\ybm)]_{i_1,\ldots,i_k}.
\end{align}
The last term is also simplified using Eq.~(\ref{eq:TweedieforPoisson_multi}),
which yields
\begin{equation}\label{eq:res3}
    [\mubm_{k}(\ybm)]_{i_1, \cdots,i_k} \Big([\mubm_1(\ybm)]_{i_{k+1}}  - \psi(\ybm_{i_{k+1}}+1) \Big), 
\end{equation}
where in the last equality, we used the fact that $\etabm_{i_{k+1}}(\xbm) = [\mubm_1(\ybm)]_{i_{k+1}}$.
By combining Eq.~(\ref{eq:res1_compact}),~(\ref{eq:res2_compact}), and~(\ref{eq:res3}), we have
\begin{align}
   \nonumber &\frac{ \partial [\mubm_k(\ybm)]_{i_1, \cdots, i_k}}{\partial \ybm_{i_{k+1}}} \\
   \nonumber& =-\sum_{j = 1}^{k}   \nabla_{\ybm_{i_{k+1}}} [\mubm_1(\ybm)]_{i_j}  [\mubm_{k-1}(\ybm)]_{\ell_j} +[\mubm_{k+1}(\ybm)]_{i_1, \cdots, i_{k+1}} \\
   \nonumber &+ \Big (  [\mubm_1(\ybm)]_{i_{k+1}} - \psi(\ybm_{i_{k+1}}+1)\Big )[\mubm_{k}(\ybm)]_{i_1, \cdots,i_k} \\
   \nonumber&- \Big([\mubm_1(\ybm)]_{i_{k+1}} - \psi(\ybm_{i_{k+1}}+1) \Big)[\mubm_{k}(\ybm)]_{i_1, \cdots,i_k} \\
   \nonumber& = -\sum_{j = 1}^{k}   \nabla_{\ybm_{i_{k+1}}} [\mubm_1(\ybm)]_{i_j}  [\mubm_{k-1}(\ybm)]_{\ell_j}+[\mubm_{k+1}(\ybm)]_{i_1, \cdots,i_{k+1}}. 
\end{align}
Putting all the results together gives the desired results.

\end{proof}

\end{document}